\documentstyle[aps,prb]{revtex}

\begin{document}
\draft
\title{Open $su(4)$-invariant spin ladder with boundary defects}
\author{Yupeng Wang}
\address{Department of Physics, Florida State University, Tallahassee, 
FL 32306, USA\\
and Institute of Physics, Chinese Academy of Sciences, \\ 
Beijing 100080, People's Republic of China}
\author{P. Schlottmann}
\address{Department of Physics, Florida State University, Tallahassee, 
FL 32306, USA}
\maketitle
\begin{abstract}
The integrable $su(4)$-invariant spin-ladder model with boundary defect 
is studied using the Bethe ansatz method. The exact phase diagram for
the ground state is given and the boundary quantum critical behavior is  
discussed. It consists of a gapped phase in which the rungs of the ladder
form singlet states and a gapless Luttinger liquid phase. It is found 
that in the gapped phase the boundary bound state corresponds to an 
unscreened local moment, while in the Luttinger liquid phase the local 
moment is screened at low temperatures in analogy to the Kondo effect.
\end{abstract}

\section{Introduction}

Spin ladder systems are an active field of research in condensed 
matter physics experimentally realized in some quasi-one-dimensional 
materials. \cite{1} The simplest example is the two-leg isotropic 
spin-$1/2$ ladder which has a gapped ground state. Generalizations  
of ladders to more legs and couplings beyond nearest neighbor 
exchanges show a remarkably rich behavior and can interpolate 
among a variety of systems. \cite{2,3,4} For instance a dimerized phase 
driven by biquadratic interactions was predicted in Ref. [2] and then
demonstrated in a generalized spin-ladder model\cite{3} by constructing 
the exact ground state and studying elementary excitations. Recently, 
a modulation structure induced by frustration was reported in a two-leg 
spin ladder. \cite{5} Quantum phase transitions from a gapped phase 
to a gapless phase were theoretically predicted and experimentally 
studied in the Heisenberg ladder system $Cu_2(C_5H_{12}N_2)_2Cl_4$ 
in the presence of a magnetic field. \cite{6}

Solutions of integrable models provide a useful starting point for 
the understanding of more general correlated many-body systems. A 
few integrable spin-ladder models have recently been proposed.
\cite{7,8,9,10} For instance the integrable $su(4)$-invariant spin 
ladder \cite{7} represents a special case of the Nesesyan-Tsvelik model.
\cite{2} This model was recently generalized to the multi-leg case 
\cite{11} and hole-doping of the ladder was also studied. \cite{12} 
In general some four-spin interaction terms must be included in 
integrable models (required by the integrability), which could be 
either related to spin-phonon mediated interactions or in the 
hole-doped phase they could be generated by the Coulomb repulsion 
between the holes moving in the spin correlated background. \cite{2} 
The importance of a biquadratic interaction for properties of 
$CuO_2$ plaquettes was pointed out in Ref. [13] and its effect on 
excitations in a spin ladder was studied in Ref. [3]. Some experiments 
indeed revealed that such multi-spin interactions are realized in 
solid films of $^3He$ absorbed on graphite, \cite{14} in a 
two-dimensional Wigner solid of electrons formed in a $Si$ inversion 
layer, \cite{15} in $bcc$ solid $^3He$, \cite{16} and in heavy 
fermion systems.

Impurities always play a relevant role in low-dimensional systems. 
This is especially the case in Luttinger liquids, where an impurity
may drive the system to a strong coupling fixed point, \cite{Kane}
which corresponds to an open boundary condition at the impurity 
site for low energy excitations. Therefore, the boundary impurity
is of particular interest in quasi one-dimensional systems. Typical
examples are the spin chain with boundary magnetic fields, 
\cite{19,Alcaraz} or equivalently two-dimensional classical 
statistical systems with boundary fields, \cite{Alcaraz,21} and 
a quantum impurity coupled to a one-dimensional strongly correlated 
electron host. \cite{20,25} 

In this paper we study the open $su(4)$-invariant spin ladder with a 
boundary defect. Boundary effects can arise for example if (i) the 
transverse coupling at the boundary is different from that in the 
bulk or if (ii) the rung-rung coupling at the boundary is different 
from that in the bulk. We will consider the following two model 
Hamiltonians:
\par\noindent
{\bf Model I}
\begin{equation}
H=\frac14\sum_{j=1}^{N-1}(1+\vec{\sigma}_j\cdot\vec{\sigma}_{j+1})
(1+\vec{\tau}_j\cdot\vec{\tau}_{j+1})
+\frac14J\sum_{j=2}^N\vec{\sigma}_j\cdot\vec{\tau}_j
+\frac14J'\vec{\sigma}_1\cdot\vec{\tau}_1,
\end{equation}
\par\noindent
{\bf Model II}
\begin{eqnarray}
H &=& \frac14\sum_{j=2}^{N-1}(1+\vec{\sigma}_j\cdot\vec{\sigma}_{j+1})
(1+\vec{\tau}_j\cdot\vec{\tau}_{j+1})
+\frac14J\sum_{j=1}^N\vec{\sigma}_j\cdot\vec{\tau}_j \nonumber \\
&& \ \ \ + \frac14U(1+\vec{\sigma}_1\cdot\vec{\sigma}_2)
(1+\vec{\tau}_1\cdot\vec{\tau}_2) ,
\end{eqnarray}
where $\vec{\sigma}_j$ and $\vec{\tau}_j$ are Pauli matrices 
acting on the site $j$ of the upper leg and lower leg, respectively, 
and $J$ represents the transverse rung coupling constant in the 
bulk, while $J'$ is the coupling at the boundary rung. $U$ denotes
the rung-rung coupling strength between the first and second rungs. 
Without the boundary defects (i.e., $J'=J$, $U=1$), the model is 
exactly solvable with periodic boundary conditions. \cite{7} In
Sect. II we show that model I is integrable for arbitrary $J'$, 
obtain the corresponding Bethe ansatz solution, and discuss the
physical consequences of the boundary bound state. In Sect. III
the exact solution of model II is presented for arbitrary $U$. 
Conclusions follow in Sect. IV.

\section{Solution of model I}

\subsection{Bethe ansatz formulation}

The quantum states of a single rung are $|\sigma_j^z,\tau_j^z>$.
It is convenient to define
\begin{eqnarray}
|0>=\frac1{\sqrt2}(|\uparrow,\downarrow>-|\downarrow,\uparrow>), 
\nonumber\\
|1>=|\uparrow,\uparrow>,{~~~~}|3>=|\downarrow,\downarrow> , \\
|2>=\frac1{\sqrt2}(|\uparrow,\downarrow>+|\downarrow,\uparrow>), 
\nonumber
\end{eqnarray}
which satisfy the orthogonality relation $<\alpha|\beta>=0$. For
simplicity we have omitted the rung index $j$ and obviously, the 
first state denotes a singlet rung and the latter three the triplet 
states. We introduce the Hubbard operators
\begin{eqnarray}
X^{\alpha\beta}=|\alpha><\beta|, {~~~~~~~~}\alpha,\beta=0,1,2,3,
\end{eqnarray}
and rewrite Hamiltonian (1) as
\begin{eqnarray}
H=\sum_{j=1}^{N-1}\sum_{\alpha,\beta=0}^3X_j^{\alpha\beta}
X^{\beta\alpha}_{j+1}-J\sum_{j=2}^NX^{00}_j-J'X^{00}_1+\frac14J
(N-1)+\frac14J' ,
\end{eqnarray}
and the total number of $\alpha$ rungs can be expressed as 
$N_\alpha = \sum_{j=1}^N X_j^{\alpha\alpha}$. In this way we have 
reduced model I to an $su(4)$-invariant spin chain with an 
effective magnetic field or, equivalently, to an $su(4)$-invariant 
$t-J$ model \cite{17,18} with a finite chemical potential and a 
boundary potential. Both the effective chemical potential $J$ 
and the boundary field $J'$ lift the $su(4)$ symmetry of the 
Hamiltonian. The Hamiltonian (5) can be diagonalized via either 
the algebraic Bethe ansatz \cite{19,21,20} or the coordinate Bethe 
ansatz. \cite{22} As the pseudo vacuum we chose the state in which 
all rungs are in a singlet, 
\begin{eqnarray}
|\Omega>\equiv|0_1>\otimes|0_2>\otimes\cdots\otimes|0_N>.
\end{eqnarray}
The Bethe wave functions can be constructed as
\begin{eqnarray}
|\Psi>=\sum_{\{j_m,\alpha_m\}}\Psi_{\alpha_1,\cdots,\alpha_{M_1}}
(j_1,\cdots,j_{M_1}) \times X_{j_1}^{\alpha_1 0} \cdots 
X_{j_{M_1}}^{\alpha_{M_1} 0}|\Omega>,
\end{eqnarray}
where $\Psi$ is an amplitude, the sum over $j_m$ runs from $1$ to 
$N$ and the sum over $\alpha_m$ from $1$ to $3$. The elimination of
the ``unwanted" terms from $H |\Psi>$ yields the standard nested 
Bethe ansatz equations (BAE):
\begin{eqnarray}
\left(\frac{\lambda_j-\frac i2}{\lambda_j+\frac i2}\right)^{2N}=
-\frac{\lambda_j-\frac i2\eta}{\lambda_j+\frac i2\eta}
\prod_{r=\pm}\prod_{l\neq j}^{M_1}\frac{\lambda_j-r\lambda_l
-i}{\lambda_j-r\lambda_l+i}\prod_{\alpha=1}^{M_2}
\frac{\lambda_j-r\mu_\alpha+\frac i2}{\lambda_j-r\mu_\alpha
-\frac i2},\nonumber\\
\prod_{r=\pm}\prod_{\beta\neq \alpha}^{M_2}
\frac{\mu_\alpha-r\mu_\beta-i}{\mu_\alpha-r\mu_\beta+i}=\prod_{r=\pm}
\prod_{j=1}^{M_1}
\frac{\mu_\alpha-r\lambda_j-\frac i2}{\mu_\alpha-r\lambda_j+\frac i2}
\prod_{\delta=1}^{M_3}
\frac{\mu_\alpha-r\nu_\delta-\frac i2}
{\mu_\alpha-r\nu_\delta+\frac i2},\\
\prod_{r=\pm}\prod_{\gamma\neq\delta}^{M_3}
\frac{\nu_\delta-r\nu_\gamma-i}
{\nu_\delta-r\nu_\gamma+i}=\prod_{r=\pm}
\prod_{\alpha=1}^{M_2}
\frac{\nu_\delta-r\mu_\alpha-\frac i2}
{\nu_\delta-r\mu_\alpha+\frac i2},\nonumber
\end{eqnarray}
where the parameter $\eta$ is determined by $J$ and $J'$ via
\begin{eqnarray}
J-J'-1=\frac{1+\eta}{1-\eta},
\end{eqnarray}
and $M_1=N_1+N_2+N_3$, $M_2=N_2+N_3$ and $M_3=N_3$. Here $\lambda_j$, 
$\mu_\alpha$ and $\nu_\delta$ represent the rapidities of the flavor 
waves. The energy spectrum of the Hamiltonian (1) is given by
\begin{eqnarray}
E=-\sum_{j=1}^{M_1}(\frac 1{\lambda_j^2+\frac14}-J)-\frac34JN-
\frac34(J'-J)+N-1.
\end{eqnarray}

\subsection{Ground state properties}

For the periodic boundary conditions all the rapidities are real 
in the ground state. For $J=4$ the system has a quantum critical 
point. \cite{7} When $J>4$, the ground state is the pseudo vacuum
$|\Omega>$, i.e. all rungs are in the singlet state. For $J<4$, on
the other hand, there is a continuum of excitations given by a
Luttinger liquid of in general three components. 

The boundary defect may change the phase diagram close to the end
point of the ladder. In particular, imaginary solutions of the BAE
arising from the boundary scattering matrix (first factor on the
right-hand side of the first set of Eqs.(8)) correspond to wave
functions that fall off exponentially from the boundary. In fact, 
$\lambda=i\eta/2$ is always a solution of the BAE in the 
thermodynamic limit $N\to\infty$ for $\eta>0$ and $\eta\neq1$ 
($\eta=1$ is a singular point of Eq.(9) which corresponds 
to $J'=\pm\infty$). This imaginary mode represents the boundary 
bound state corresponding to a triplet rung. A careful analysis of
the energy carried by the imaginary mode yields that the boundary 
bound state is not always stable (occupied) in the ground state. 
We limit ourselves to the situation of antiferromagnetic coupling
($J>0$). From Eq.(10) we have that the energy of the imaginary 
mode is
\begin{eqnarray}
\epsilon_b=J-\frac4{1-\eta^2}.
\end{eqnarray}
The boundary bound state is stable if $\epsilon_b<0$. Otherwise 
the imaginary mode represents an excited state. We have to 
distinguish the cases $J > 4$ from $J < 4$. (i) For $J>4$, 
there is a critical line given by
\begin{eqnarray}
\eta_c=\sqrt{1-\frac4J},
\end{eqnarray}
which separates the spin singlet rung ground state ($0<\eta<\eta_c$
and $\eta > 1$) from the spin triplet ground state ($\eta_c<\eta<1$) 
at the boundary of the ladder. For $0<\eta<\eta_c$ and $\eta > 1$ 
we have $\epsilon_b>0$, while for $\eta_c<\eta<1$ $\epsilon_b$ is 
negative and the boundary bound state is filled. (ii) For $0<J<4$ 
the bulk corresponds to a three component Luttinger liquid with 
real $\lambda$, $\mu$ and $\nu$ modes. Here the boundary bound 
state is stable in the whole parameter region $0<\eta<1$, but it is
empty if $\eta > 1$. The boundary triplet state for $0<\eta<1$ is 
coupled to the continuum giving rise to a Kondo-like screening.

Consider now the response of the system to an external magnetic 
field $h$. The magnetic field couples to the ladder via the Zeeman 
effect, i.e. the Hamiltonian has an extra term $-h\sum_{j=1}^N
(X_j^{11}-X_j^{33})$. The critical line $J = 4$ separating the 
gapped and gapless regions is now shifted to $J - h =4$. The 
bound state energy is also reduced by $h$. Hence, for $J-h>4$, 
the critical line Eq.(12) is now given by $\eta_c=
\sqrt{1-4/(J-h)}$. The threefold degeneracy of the triplet rung
state at the boundary is lifted by the field, so that an 
arbitrarily small field induces a finite magnetization $\pm1$ 
(depending on the direction of the field) due to the stabilization 
of the bound state. Therefore, the boundary quantum phase transition 
at $\eta=\eta_c$ is of first order and the susceptibility is 
divergent at $T=0$, following a Curie law. 

To summarize, the boundary phase diagram for $J>h$ is shown in 
Fig. 1(a) and consists of seven regions. For $J-h>4$ the bulk 
is gapped and we have argued that for $\eta<0$ there is no bound 
state and the rungs all form singlets. For $0<\eta<\eta_c$ the 
bound state is not stable and all rungs are in singlet states. 
For $\eta_c<\eta<1$ there is a triplet state at the boundary
(the triplet wavefunction falls off exponentially into the bulk 
of the ladder), and finally for $\eta>1$ the bound state is 
again unstable. For $J-h<4$ the bulk is a Luttinger liquid, 
without a bound state for $\eta <0$, with a stable bound 
state for $0<\eta<1$ and with an unstable bound state for 
$\eta >1$. Below we show that a Kondo-like screening occurs 
for $0<\eta<1$.

It is also interesting to study the situation for $h-J>0$. The
bulk is then always a Luttinger liquid and only four cases
for the boundary bound state have to be distinguished (see
Fig. 1(b)). We assume here that $h-J$ is sufficiently small
so that the spin ladder is not spin polarized. If $\eta <0$ 
there is no bound state, for $0<\eta<1$ the bound state
is filled with a predominantly spin-up triplet state (the 
Kondo screening is quenched by the magnetic field), for 
$1<\eta<\eta_c$ there is an empty bound state, and for 
$\eta>\eta_c$ the bound state is again stable (with 
magnetic field quenched Kondo screening).  

The thermodynamics of the boundary defect can also be derived
from the BAE, Eq.(8), following the standard method. \cite{23,24} 
The thermodynamic BAE allow us to study the boundary quantum 
critical behavior. The boundary defect induces a ``ghost spin'' 
$\eta$. However, unlike in $su(2)$-invariant models, \cite{25}
the ghost spin does not lead to an anomalous remnant entropy 
because the $su(4)$ symmetry in the present model is already
lifted by the finite $J$.  

\subsection{The quantum critical line $J=4$}

Along the quantum critical line $J=4$, the boundary defect can 
show critical behavior as the bulk does. \cite{7} We now consider 
the case $\eta>1$. In zero magnetic field, the ground state of
the bulk consists only of singlet rungs. In a weak magnetic field 
some triplet rungs with $S^z=1$ appear in the ground state, 
while $N_2$ and $N_3$ still remain equal to zero, since the energy 
of the state $|2>$ is unchanged and that of the state $|3>$ is 
increased ($h>0$). We denote with $\rho(\lambda)$ the distribution 
of real $\lambda$ modes, including the boundary-defect contribution. 
>From the BAE, Eq.(12), we obtain
\begin{eqnarray}
\rho(\lambda)+\int_{-\Lambda}^\Lambda d\lambda' a_2(\lambda-\lambda')
\rho(\lambda')= a_1(\lambda) - \frac{1}{2N} a_\eta(\lambda),
\end{eqnarray}
where $a_n(\lambda)=n/2\pi[\lambda^2+(n/2)^2]$ and $\Lambda^2=
1/(4-h)-1/4$. For $h<<1$, we have $\Lambda\approx\sqrt{h}/4$ and 
Eq.(13) can be solved by iteration,
\begin{eqnarray}
\rho(\lambda)=\left(\frac{2}{\pi}-\frac{1}{\pi\eta N}\right)
\left(1 - \frac{2\Lambda}{\pi} \right) + \cdots \ ,
\end{eqnarray}
with the ground state energy given by
\begin{eqnarray}
E=\int_{-\Lambda}^\Lambda d\lambda \left(4-\frac1{\lambda^2+\frac14}
-h\right)\rho(\lambda) -\frac34JN-\frac34(J'-J)+N-1 \ .
\end{eqnarray}
Combining Eqs.(14) and (15) we obtain the susceptibility of the 
system
\begin{eqnarray}
\chi =  -\frac{\partial^2 E}{\partial h^2} = 
\left(\frac{2}{\pi}-\frac{1}{\pi\eta N}\right)
\left(\frac{1}{4}h^{-\frac12} - \frac{1}{3\pi}\right)+O(h^{\frac12}).
\end{eqnarray}
The susceptibility diverges with the square root of the field as a 
consequence of the van Hove singularity of the empty $\lambda$ band.
The boundary bound state removes one degree of freedom from the 
bulk, so that its contribution to the susceptibility is negative.
This result is not surprising, since in this case $J'$ is much 
larger than $J$. Consequently the boundary rung is in a tight 
singlet and hence insensitive to the field, so that the whole 
susceptibility is reduced. 

Similar arguments indeed yield a positive boundary susceptibility 
for $\eta<0$. As discussed before, in the region $0<\eta<1$, a 
stable boundary bound state occurs and a small field already 
induces a finite magnetization. 

With a simple scaling approach we find that the low temperature 
specific heat and the magnetic susceptibility of the boundary 
defect at the line $J =4$ behaves as
\begin{eqnarray}
\delta C(T)\sim T^{\frac12}, {~~~~~~~~~~}\delta \chi(T)
\sim T^{-\frac12}.
\end{eqnarray}
Such a result can also be predicted by a simple spin wave theory 
with a dispersion relation $\epsilon(k)\sim k^2$ or alternatively
exactly via the low temperature expansion of the thermodynamic 
BAE. \cite{26,27} The boundary critical exponents in Eqs.(16-17) 
are exactly the same as for the bulk. \cite{7}

\subsection{Kondo effect in the gapless phase}

In the gapless phase, $0<J<4$, the system is a three component 
Luttinger liquid. In the sense of the $su(4)$ $t-J$ model, the 
triplet rungs are considered spin-1 hard-core bosons. The $\lambda$ 
rapidities represent the charge sector, while the $\mu$ and $\nu$
rapidities parametrize the spin degrees of freedom, which have 
$su(3)$ invariance. 

As discussed above (see Fig. 1(a)) a stable boundary bound state 
only exists at low temperatures for $0<\eta<1$. The boundary bound 
state corresponds to a local moment with spin 1. The boundary
coupling $J'$ does not break the $su(3)$-invariance of the hard-core 
bosons and the boundary local moment is spin compensated in analogy
to the Kondo effect. 

To show this we explicitly consider the imaginary mode $i\eta/2$ in 
the BAE, which then become
\begin{eqnarray}
\left(\frac{\lambda_j-\frac i2}{\lambda_j+\frac i2}\right)^{2N}
&=& -\frac{\lambda_j-\frac i2\eta} {\lambda_j+\frac i2\eta}
\frac{\lambda_j-i(1-\frac\eta2)}{\lambda_j+i(1-\frac\eta2)}
\frac{\lambda_j-i(1+\frac\eta2)}{\lambda_j+i(1+\frac\eta2)}
\nonumber\\
&\times& \prod_{r=\pm}\prod_{l\neq j}^{M_1-1}\frac{\lambda_j-
r\lambda_l-i}{\lambda_j-r\lambda_l+i}\prod_{\alpha=1}^{M_2}
\frac{\lambda_j-r\mu_\alpha+\frac i2}{\lambda_j-r\mu_\alpha
-\frac i2}, \nonumber\\
\prod_{r=\pm}\prod_{\beta\neq \alpha}^{M_2}\frac{\mu_\alpha
-r\mu_\beta-i}{\mu_\alpha-r\mu_\beta+i}
&=& \prod_{r=\pm} \frac{\mu_\alpha-\frac i2(1+r\eta)}
{\mu_\alpha+\frac i2(1+r\eta)}\prod_{j=1}^{M_1-1}
\frac{\mu_\alpha-r\lambda_j-\frac i2}{\mu_\alpha-r\lambda_j
+\frac i2} \prod_{\delta=1}^{M_3} \frac{\mu_\alpha-r\nu_\delta
-\frac i2}{\mu_\alpha-r\nu_\delta+\frac i2}, \\
\prod_{r=\pm}\prod_{\gamma\neq\delta}^{M_3}\frac{\nu_\delta
-r\nu_\gamma-i}{\nu_\delta-r\nu_\gamma+i}
&=& \prod_{r=\pm} \prod_{\alpha=1}^{M_2} \frac{\nu_\delta-
r\mu_\alpha-\frac i2}{\nu_\delta-r\mu_\alpha+\frac i2} \ .
\nonumber
\end{eqnarray}
For large $N$ the solutions of the BAE, Eq. (18), are strings of 
arbitrary length for all three sets of rapidities. We introduce the 
usual densities of $\lambda$, $\mu$, $\nu$ strings 
$\rho_{1,n}(\lambda)$, $\rho_{2,n}(\mu)$, $\rho_{3,n}(\nu)$, and 
their respective hole densities $\rho_{1,n}^h(\lambda)$, 
$\rho_{2,n}^h(\mu)$, $\rho_{3,n}^h(\nu)$. In the thermodynamic limit
these densities satisfy the following integral equations \cite{28}
\begin{eqnarray}
\rho_{1,n}^h(\lambda) &+& \sum_m A_{mn} \rho_{1,m}(\lambda)
= \sum_m B_{mn} \rho_{2,m}(\lambda) + a_n(\lambda) ,\nonumber \\
\rho_{2,n}^h(\lambda) &+& \sum_m A_{mn} \rho_{2,m}(\lambda)
= \sum_m B_{mn}(\rho_{1,m}(\lambda) + \rho_{3,m}(\lambda)) , \\
\rho_{3,n}^h(\lambda) &+& \sum_m A_{mn} \rho_{3,m}(\lambda)
= \sum_m B_{mn} \rho_{2,m}(\lambda) , \nonumber
\end{eqnarray}
where we neglected the boundary driving terms which are of order 
$N^{-1}$. Here
\begin{eqnarray}
A_{mn} &=& [m+n]+2[m+n-2]+\cdots+2[|m-n|+2]+[|m-n|] \ , \nonumber \\
B_{mn} &=& \sum_{l=1}^{min\{m,n\}} [m+n-2l+1] , \nonumber 
\end{eqnarray}
and $[n]$ is the integral operator with kernel $a_n(\lambda)$ and
$a_0(\lambda)$ is the $\delta$-function.

The free energy functional is given by
\begin{eqnarray}
F/N &=& \sum_{r,n} \int d\lambda \{ \epsilon_{r,n}\rho_{r,n}(\lambda)
- T [\rho_{r,n}(\lambda)+\rho_{r,n}^h(\lambda)] \ln[\rho_{r,n}(\lambda)
+\rho_{r,n}^h(\lambda)] \nonumber \\
&+& T \rho_{r,n}(\lambda) \ln\rho_{r,n}(\lambda) + T 
\rho_{r,n}^h(\lambda) \ln\rho_{r,n}^h(\lambda) \} \ ,
\end{eqnarray}
where $\epsilon_{1,n} = -2\pi a_n(\lambda)+n(J-h)$, $\epsilon_{2,n} = 
\epsilon_{3,n}=nh$. Minimizing Eq. (20) with respect to the densities 
and taking into account the relations (19), we obtain \cite{28,29}
\begin{eqnarray}
\ln(1+\eta_{r,n})=\frac{\epsilon_{r,n}}T+\sum_{m,s}A_{mn}^{rs}
\ln(1+\eta_{s,m}^{-1}), {~~~~~}r,s=1,2,3
\end{eqnarray}
with $A_{mn}^{rs}=A_{mn}\delta_{r,s}-B_{mn}(\delta_{r,s+1}+
\delta_{r,s-1})$, $\eta_{r,n}=\rho_{r,n}^h/\rho_{r,n}$, and 
$\eta_{0,n}^{-1}=\eta_{4,n}^{-1}\equiv 0$. An equivalent set of 
integral equations is 
\begin{eqnarray}
\ln\eta_{r,n} &=& G \star [\ln(1+\eta_{r,n+1}) + \ln(1+\eta_{r,n-1})]
- G \star [\ln(1+\eta_{r+1,n}^{-1}) + \ln(1+\eta_{r-1,n}^{-1})] \ ,
\nonumber \\
\ln\eta_{r,1} &=& -\frac{2\pi}{T} G(\lambda) \delta_{r,1} + G
\star \ln(1+\eta_{r,2}) - G \star [\ln(1+\eta_{r+1,1}^{-1}) +
\ln(1+\eta_{r-1,1}^{-1})] \ , \\
&& \lim_{n\to\infty} \frac{\ln\eta_{1,n}}n = \frac{J-h}{T} \ \ , \
\lim_{n\to\infty}\frac{\ln\eta_{2,n}}n = \lim_{n\to\infty}
\frac{\ln\eta_{3,n}}n = \frac{h}{T} \equiv 2x_0 \ \ , \nonumber
\end{eqnarray}
where $\star$ denotes convolution and $G(\lambda) = [2\cosh(\pi 
\lambda)]^{-1}$. The equilibrium free energy is
\begin{eqnarray}
F/N = - T \sum_n [n] \ln(1+\eta_{1,n}^{-1}) \ . \nonumber
\end{eqnarray}

At low $T$ the exchange $J$ gives rise to a Fermi surface for the
charges, which are only significantly populated in the interval
$|\lambda|<\Lambda=\sqrt{1/J-1/4}$, but are unoccupied for 
$|\lambda|>\Lambda$. The low energy spin excitations, on the
other hand, take place at very large rapidities (for $h = 0$ the
spin Fermi surface is at $\infty$). Hence, at low $T$ the charge
and spin sectors are well separated and only weakly coupled. 
Assuming complete decoupling of the spin and charge sectors,
a solution of Eq. (22) can be easily obtained for large $n$
\begin{eqnarray}
\eta_{2,n} = \eta_{3,n} = \frac{\sinh(nx_0)\sinh(n+1)x_0}{\sinh 
x_0\sinh(2x_0)} - 1 \ .
\end{eqnarray}

Although the boundary bound state has also some charge fluctuations  
(triplet-singlet rung admixture), these do not affect the dynamics
of the spin (weak coupling of spin and charge sectors). We can then 
limit ourselves to the spin degrees of freedom of the bound state. 
Its contribution to the free energy is
\begin{eqnarray}
F_{bs} = -\frac12 T \sum_{n=1}^\infty \int d\lambda 
[a_n(\lambda-\frac i2\eta) + a_n(\lambda+\frac i2\eta)]
\ln[1+\eta_{2,n}^{-1}] 
\end{eqnarray}
or, equivalently, 
\begin{eqnarray}
F_{bs}=F_{bs}^0 - \frac12 T \sum_{r=\pm} \sum_{q=1}^2 \int d\lambda
G_q(\lambda+ir\eta/2) \ln[1+\eta_{q+1,1}] \ ,
\end{eqnarray}
where $F_{bs}^0$ the ground state energy of the local moment and
\begin{eqnarray}
G_q(\lambda) = \frac{\sin[\pi(1-q/3)]}{\cosh(2\pi\lambda/3)+
\cos[\pi(1-q/3)]} \ . \nonumber
\end{eqnarray}
Substituting Eq.(23) into Eq.(25) we readily obtain that the residual 
entropy of the magnetic moment is exactly zero, which means that the
boundary bound state is fully screened by the Kondo effect. These 
results are analogous to those of a Coqblin-Schrieffer impurity 
embedded into the same host. \cite{30} 

\section{Solution of Model II}

To show that model II is integrable we rewrite the Hamiltonian (2) as
\begin{eqnarray}
H=H_0+H_1,\nonumber\\
H_0=\sum_{j=2}^{N-1}P_{jj+1}+UP_{12},\\
H_1=-J\sum_{j=1}^NX_j^{00}+\frac14NJ,\nonumber
\end{eqnarray}
where $P_{ij}=(1+\vec{\sigma}_i\cdot\vec{\sigma}_j)(1+\vec{\tau}_i
\cdot\vec{\tau}_j)/4$ is the permutation operator between rung $i$ 
and rung $j$. Obviously, $[H_0,H_1]=0$, which means that they can 
be diagonalized simultaneously. We define the Lax operators
\begin{eqnarray}
S_{ij}(\lambda)\equiv \lambda-iP_{ij}
\end{eqnarray}
which satisfy the Yang-Baxter relations \cite{31}
\begin{eqnarray}
S_{ij}(\lambda-\mu)S_{ik}(\lambda)S_{jk}(\mu) 
= S_{jk}(\mu)S_{ik}(\lambda)S_{ij}(\lambda-\mu) .
\end{eqnarray}
As shown in Refs.[18,22], the monodromy matrix
\begin{eqnarray}
T_\tau(\lambda)\equiv S_{N\tau}(\lambda)S_{N-1\tau}(\lambda)
\cdots S_{2\tau}(\lambda)S_{1\tau}(\lambda-ic)S_{1\tau}(\lambda+ic)
S_{2\tau}(\lambda)\cdots S_{N-1\tau}(\lambda)S_{N\tau}(\lambda)
\end{eqnarray}
satisfies the reflection Yang-Baxter equation
\begin{eqnarray}
S_{\tau\tau'}(\lambda-\mu)T_\tau(\lambda)S_{\tau\tau'}(\lambda+\mu)
T_{\tau'}(\mu)=T_{\tau'}(\mu)S_{\tau\tau'}(\lambda+\mu)
T_\tau(\lambda)S_{\tau\tau'}(\lambda-\mu),
\end{eqnarray}
where $c$ is an arbitrary constant and $\tau$ and $\tau'$ are 
indices of the 4-dimensional auxiliary space. From Eq.(30) we have
\begin{eqnarray}
[t(\lambda),t(\mu)]=0
\end{eqnarray}
with $t(\lambda)\equiv tr_\tau T_\tau(\lambda)$. Therefore, 
$t(\lambda)$ serves as the generating functional of a variety of 
conserved quantities. $H_0$ is related to $t(\lambda)$ as
\begin{eqnarray}
H_0=\frac i{8(1-c^2)}(-1)^{N+1}\frac{dt(\lambda)}{d\lambda}|_{\lambda=0},
\end{eqnarray}
provided that $U=1/(1-c^2)$. Hence, $H_0$ and Hamiltonian (2) are 
integrable in the sense of the algebraic Bethe ansatz. Following the
standard procedure we readily obtain the BAE of model II 
\begin{eqnarray}
\left(\frac{\lambda_j-\frac i2}{\lambda_j+\frac i2}\right)^{2(N-1)}=
\frac{\lambda_j+i(c+\frac12)}{\lambda_j-i(c+\frac12)}
\frac{\lambda_j-i(c-\frac12)}{\lambda_j+i(c-\frac12)}
\prod_{r=\pm}\prod_{l\neq j}^{M_1} \frac{\lambda_j-r\lambda_l-i}{
\lambda_j-r\lambda_l+i}\prod_{\alpha=1}^{M_2}
\frac{\lambda_j-r\mu_\alpha+\frac i2}{\lambda_j-r\mu_\alpha
-\frac i2} , \nonumber \\
\prod_{r=\pm}\prod_{\beta\neq \alpha}^{M_2}\frac{\mu_\alpha
-r\mu_\beta-i}{\mu_\alpha-r\mu_\beta+i}=\prod_{r=\pm} 
\prod_{j=1}^{M_1} \frac{\mu_\alpha-r\lambda_j-\frac i2}
{\mu_\alpha-r\lambda_j+\frac i2} \prod_{\delta=1}^{M_3}
\frac{\mu_\alpha-r\nu_\delta-\frac i2}{\mu_\alpha-r\nu_\delta
+\frac i2} , \\
\prod_{r=\pm}\prod_{\gamma\neq\delta}^{M_3}\frac{\nu_\delta
-r\nu_\gamma-i}{\nu_\delta-r\nu_\gamma+i}=\prod_{r=\pm}
\prod_{\alpha=1}^{M_2} \frac{\nu_\delta-r\mu_\alpha-\frac i2}
{\nu_\delta-r\mu_\alpha+\frac i2} \nonumber
\end{eqnarray}
with the energy eigenvalues given by
\begin{eqnarray}
E=-\sum_{j=1}^{M_1}(\frac 1{\lambda_j^2+\frac14}-J)-\frac34JN+U+N-2.
\end{eqnarray}

The Hamiltonian (32) is only Hermitian if the parameter $c$ is 
either real or imaginary. For imaginary $c$, $U<1$, and the first 
rung is weakly coupled to the bulk. For real $c$ (we can consider 
$c>0$ because $\pm c$ yields the same $U$) the imaginary mode 
$\lambda_b=i(c-1/2)$ is a solution of the BAE, Eq.(33), in the 
thermodynamic limit $N\to\infty$ for $c>1/2$. This bound state
again corresponds to a triplet boundary bound state, i.e. its wave 
function falls off exponentially with the distance from the boundary. 
The energy of the bound state is $\epsilon_b = J - 1/(c-c^2)$, which 
can be stable only if $\epsilon_b < 0$, i.e. in the region $1/2<c<1$. 
This corresponds to $U > 1$ for which the first rung is strongly 
coupled to the bulk. For $c > 1$, i.e. $U < 0$, the bound state has 
positive energy and is empty. The first rung is then ferromagnetically 
coupled to the ladder.   

As in the case of model I, there is a critical line for $J>4$ 
given by
\begin{eqnarray}
J_c=\frac1{(c-c^2)} \ ,
\end{eqnarray}
which separates the region of an occupied (stable) and empty (unstable) 
bound state. Hence, when $J>J_c$ the ground state is a spin singlet, 
while for $4<J<J_c$ and $1/2<c<1$ the boundary bound state is stable in
the ground state. 

In summary, the physical properties of the two models are very similar. 
The main difference can be understood in terms of the number of ``ghost
spin'' solutions (number of bound state solutions) of the BAE, \cite{25}
which correspond to images of the real local moment. In the Kondo regime 
for $J < 4$ there are two ``ghost spins'', $c+1/2$ and $c-1/2$, in the 
spin sector of the second model, while there is only one in model I. 
This can be read off from the impurity factors in the BAEs (8) and (33).
Since the effects of the ghost spin contributions are additive, the 
physics in both situations is very similar. 

\section{Conclusions}

We studied two models for boundary defects of the open two-leg 
$su(4)$-invariant spin ladder. In model I the transverse coupling 
at the boundary rung is different from the bulk, while in model II
the coupling of the first rung to the ladder is different from the
rung-rung coupling in the bulk. The two models under consideration 
are integrable and we obtained the exact solution by means of 
Bethe's ansatz. Depending on the model parameters three situations
may arise in the thermodynamic limit: (i) there is no imaginary 
mode solution of the BAE and hence the states of the first rung
are part of the continuum of the bulk, (ii) an imaginary boundary 
mode exists, but corresponds to a positive energy $\epsilon_b$, 
i.e. the state is empty, and (iii) an imaginary boundary mode 
with negative energy exists. In case (ii) the boundary bound 
state does not affect the ground state properties, but does
contribute to the finite $T$ thermodynamics. The situation (iii)
is the most interesting one, since a spin-1 boundary bound state
is filled in the ground state. The boundary phase diagram of 
model I is shown in Figs. 1(a) and 1(b).  

The $su(4)$-invariant two-leg ladder has a critical line at 
$J = 4$. For $J > 4$ all the rungs of the bulk are in the
singlet state, while for $J < 4$ the system is a Luttinger liquid
of in general three components. For $J > 4$ a stable boundary
bound state carries a magnetic moment (triplet state with wave
function that falls off exponentially into the bulk), while 
in all other cases the boundary rung is in its singlet state. 
For $J < 4$, on the other hand, a stable boundary bound state 
carries a Kondo compensated (screened by the spin degrees of 
freedom in the Luttinger liquid) magnetic moment of spin-1, 
i.e. ultimately the ground state is a singlet. An unstable 
bound state just removes one degree of freedom from the 
Luttinger liquid. Both models considered here display similar 
properties at the boundary. 

We acknowledge the support by the National Science Foundation 
and the Department of Energy under grants No. DMR98-01751 and
No. DE-FG02-98ER45797.  Y. Wang is also supported by the 
National Science Foundation of China.

\newpage
\centerline{\bf Figure caption}
\vskip 0.2in
\noindent
{\bf Fig. 1 :} Boundary phase diagram of the two-leg $su(4)$-invariant
spin ladder with different transverse coupling at the first rung 
(model I) for (a) $J-h>0$ and (b) $h-J>0$. The interaction strength 
at the first rung is parametrized by $\eta$ defined in Eq.(9). Quantum
critical behavior with mean-field exponents is obtained along the line 
$J-h=4$. Several phases are possible at the boundary. In S all rungs
are in a singlet state and there is no boundary bound state. In S1 the
ground state consists of singlet rungs but there is an empty boundary 
bound state. TBS refers to a phase in which the boundary bound state is 
stable, i.e. there is a boundary triplet state with wave function 
falling off exponentially into the bulk. LL1, LL2 and LL3 refer to a 
Luttinger liquid with no boundary bound state, with a stable boundary 
bound state (triplet state) and with an unstable boundary 
bound state, respectively. In (b) we assumed that $h-J$ is sufficiently
small so that the ladder is not spin polarized.


\begin{references}
\bibitem{1}E. Dagotto and T.M. Rice, Science {\bf 271}, 618 (1996).
\bibitem{2}A.A. Nersesyan and A.M. Tsvelik, Phys. Rev. Lett. 
{\bf 78}, 3939 (1997).
\bibitem{3}A.K. Kolezhuk and H.-J. Mikeska, Int. J. Mod. Phys. B 
{\bf 12}, 2325 (1998); Phys. Rev. Lett. {\bf 80}, 2709 (1998);
S. Brehmer, H.-J. Mikeska, M. M\"uller, N. Nagaosa and S. Uchida, 
Phys. Rev. B {\bf 60}, 329 (1999).
\bibitem{4}I. Bose and S. Gayen, Phys. Rev. B {\bf 48}, 10653 (1993);
A. Gosh and I. Bose, Phys. Rev. B {\bf 55}, 3613 (1997).
\bibitem{5}Y. Wang, Int. J. Mod. Phys. B {\bf 13}, 3323(1999).
\bibitem{6}G. Chaboussant {\it et al.}, Phys. Rev. Lett. {\bf 80}, 
2713 (1998).
\bibitem{7}Y. Wang, Phys. Rev. B {\bf 60}, 9236 (1999).
\bibitem{8}H. Frahm and C. R\"{o}denbeck, J. Phys. A {\bf 30}, 
4467 (1997).
\bibitem{9}N. Muramoto and M. Takahashi, J. Phys. Soc. Japan 
{\bf 68}, 2098 (1999)
\bibitem{10}S. Albeverio, S.M. Fei and Y. Wang, Europhys. Lett. 
{\bf 47}, 364 (1999).
\bibitem{11}M.T. Batchelor and M. Maslen, J. Phys. A {\bf 32}, 
L337 (1999).
\bibitem{12}H. Frahm and A. Kundu, J. Phys.: Cond. Matter {\bf 11}, 
L557 (1999).
\bibitem{13}Y. Honda, Y. Kuramoto and T. Watanabe, Phys. Rev. B 
{\bf 47}, 11329 (1993).
\bibitem{14}K. Ishida, et al., Phys. Rev. Lett. {\bf 79}, 3451 
(1997); M. Roger, et al., ibid, {\bf 80}
1308 (1998).
\bibitem{15}T. Okamoto and S. Kawaji, Phys. Rev. B {\bf 57}, 9097 
(1998).
\bibitem{16}For a review, see, D.D. Osheroff, J. Low Temp. Phys. 
{\bf 87}, 297 (1992).
\bibitem{Kane}C. L. Kane and M. P. A. Fisher, Phys. Rev. Lett. 
{\bf 68}, 1220 (1992); E. S{\o}rensen, S. Eggert and I. Affleck, 
J. Phys. A {\bf 26}, 6757 (1993).
\bibitem{19}E.K. Sklyanin, J. Phys. A {\bf 21}, 2375 (1988). 
\bibitem{Alcaraz}F. C. Alcaraz, M. N. Barber, M. T. Batchelor, 
R. J. Baxter, and G. R. W. Quispel, J. Phys. A {\bf 20}, 6397
(1987).
\bibitem{21}C.M. Yung and M.T. Batchelor, Nucl. Phys. B {\bf 453}, 
552 (1995).
\bibitem{20}A. Foerster and M. Karowski, Nucl. Phys. B {\bf 396}, 
611 (1993); B {\bf 408}, 512 (1993).
\bibitem{25}Y. Wang, Phys. Rev. B {\bf 56}, 14045 (1997); Y. Wang 
and U. Eckern, Phys. Rev. B {\bf 59}, 6400 (1999); J. Dai, Y. Wang, 
and U. Eckern, Phys. Rev. B {\bf 60}, 6594 (1999); J. Dai and Y. 
Wang, Phys. Rev. B {\bf 60}, 12309 (1999).
\bibitem{17}B. Sutherland, Phys. Rev. B {\bf 12}, 3795 (1975).
\bibitem{18}P. Schlottmann, Phys. Rev. B {\bf 36}, 5177 (1987).
\bibitem{22}Y. Wang, J. Dai, Z. Hu, and F.-C. Pu, Phys. Rev. Lett. 
{\bf 79}, 1901 (1997).
\bibitem{23}C.N. Yang and C.P. Yang, J. Math. Phys. {\bf 10}, 1115 
(1969).
\bibitem{24}M. Takahashi, Prog. Theor. Phys. {\bf 46}, 401 (1971); 
{\it ibid.} {\bf 46}, 1388 (1971).
\bibitem{26}P. Schlottmann, Phys. Rev. B {\bf 33}, 4880 (1986).
\bibitem{27}Y. Wang and J. Dai, Phys. Rev. B {\bf 59}, 13561 (1999).
\bibitem{28}H. Johannesson, Phys. Lett. A {\bf 116}, 133 (1986).
\bibitem{29}N. Andrei, K. Furuya and J.H. Lowenstein, Rev. Mod. Phys. 
{\bf 55}, 331 (1983); A. M. Tsvelik and P. B. Wiegmann, Adv. Phys.
{\bf 32}, 453 (1983).
\bibitem{30}P. Schlottmann, J. Phys. Cond. Matter {\bf 10}, 2525
(1998).
\bibitem{31}C.N. Yang, Phys. Rev. Lett. {\bf 10}, 1312 (1967).
\end{references}
\end{document}